 \documentclass[a4paper,12pt]{article}

\pdfoutput=1

\RequirePackage{amsfonts,amssymb,amsbsy,amsmath}

\usepackage{graphicx}

\begin{document}

\newcommand {\tim}{n}
\newcommand {\inv}{{^{-1}}}
\newcommand {\half}{\textstyle{\frac{1}{2}}}
\newcommand {\CandLa}{{[C{\&}La]}}
\newcommand {\CandLb}{{[C{\&}Lb]}}

\newcommand {\Peak}{{\sc Peak}}

\newcommand{\bff}{{\boldsymbol f}}
\newcommand{\bfeta}{{\boldsymbol\eta}}
\newcommand{\bfdelta}{{\boldsymbol\delta}}
\newcommand{\bfdelbar}{{\overline{\boldsymbol\delta}}}
\newcommand{\bfT}{{\boldsymbol T}}
\newcommand{\bfzero}{{\boldsymbol 0}}

\newcommand{\bfDelsq}{{\boldsymbol\nabla}_\sigma^2}
\newcommand{\bfTbar}{\bar{\boldsymbol T}}
\newcommand{\bfpT}{{\boldsymbol{p}^{\mathrm{T}}}}
\newcommand{\bfPhi}{{\boldsymbol\Phi}}

\newcommand{\ssB}{{\mathsf{B}}}
\newcommand{\ssE}{{\mathsf{E}}}
\newcommand{\ssET}{{\mathsf{E}^{\rm T}}}
\newcommand{\ssG}{{\mathsf{G}}}
\newcommand{\ssH}{{\mathsf{H}}}
\newcommand{\ssI}{{\mathsf{I}}}
\newcommand{\ssLAM}{{\mathsf{\Lambda}}}
\newcommand{\ssOM}{{\mathsf{\Omega}}}
\newcommand{\ssP}{{\mathsf{P}}}
\newcommand{\ssPA}{{\mathsf{P}_A}}
\newcommand{\ssPB}{{\mathsf{P}_B}}
\newcommand{\ssPC}{{\mathsf{P}_C}}
\newcommand{\ssPD}{{\mathsf{P}_D}}

\newcommand{\dd}{{\mathrm{d}}}
\newcommand{\diag}{\mathrm{diag}}


\title{Laplace Transform Integration of a Baroclinic Model}


\author{Eoghan Harney \&\ Peter Lynch \\
        School of Mathematics \&\ Statistics, \\
        UCD, Belfield, Dublin 4, Ireland}

\maketitle

\begin{abstract}
A time integration scheme based on the Laplace Transform (LT) has
been implemented in a baroclinic primitive equation model. The LT
scheme provides an attractive alternative to the popular semi-implicit
(SI) scheme.  Analysis shows that it is more accurate than
SI for both linear and nonlinear terms of the equations. Numerical
experiments confirm the superior performance of the LT scheme. The
algorithmic complexity of the LT scheme is comparable to SI, with just
an additional transformation to vertical eigenmodes each time step, and
it provides the possibility of improving weather forecasts at comparable
computational cost.
\end{abstract}

\maketitle


\section{Introduction}
\label{sec:intro}

The accuracy and efficiency of weather and climate models has been
greatly enhanced by the introduction of better numerical algorithms
for the solution of the equations of motion. Two of the most notable
schemes are the semi-implicit (SI) scheme for treating the gravity-wave
terms and the semi-Lagrangian scheme for advection processes. Many
current models,
such as the Integrated Forecast System (IFS) of the European Centre
for Medium-Range Weather Forecasts,
use a combination of these two schemes to integrate the equations, allowing
for larger time steps and greatly reduced computational cost.

Simple analysis shows that the accuracy
of the SI scheme decreases as the time step is increased. In this
paper, we describe an alternative method of integrating the adjustment
terms. This scheme, which is based on the Laplace transform (LT),
has been implemented in a baroclinic primitive equation model, {\Peak}
(Ehrendorfer, 2012).
Analysis shows that the LT scheme is more accurate than SI for both
linear and nonlinear terms of the equations.  Numerical experiments
confirm the superior performance of the LT scheme.
Considering that the two schemes require approximately the same
computation time, the LT scheme is an attractive alternative to SI,
with the potential to improve forecasting accuracy.

The SI scheme stabilizes the high-frequency gravity wave components of
the flow, enabling the use of a large time step. However,
it also introduces phase errors, which may impact on meteorologically
significant components of the flow, thereby reducing the forecast accuracy.
The LT scheme eliminates high frequency components but maintains the
accuracy of components that are retained. This is because the time averaging 
of the linear terms in SI is not done in LT: these terms are transformed exactly.
The cut-off point for mode elimination may be chosen freely, in contrast to
the SI scheme for which there is no such means of control.

The LT scheme was first developed by Lynch as a means of balancing initial data
(Lynch, 1985a,b). An early application to integration of the equations was
described by Van Isacker and Struylaert (1986).
More recently, Clancy and Lynch (2011a) formulated an LT scheme for an Eulerian shallow water model
and demonstrated its advantages by means of a series of numerical experiments.
Clancy and Lynch (2011b) then modified the LT scheme for use in a shallow water model
with Lagrangian advection. All the above schemes used a numerical inversion of the Laplace transform. 
Lynch and Clancy (2016) showed that, with an appropriate formulation of the equations,
the inversion could be performed analytically. This greatly enhanced the computational
efficiency of the LT scheme, making it competitive with the SI method.

\subsection*{Baroclinic Forecasts}

As a preliminary step, an LT scheme has been implemented in a shallow water version
of the OpenIFS model (Carver, et al., 2018).  In the current study, we describe the
application of the LT scheme in a fully three-dimensional forecasting model, {\Peak}
(Ehrendorfer, 2012).  In \S\ref{sec:anal}, the LT and SI schemes are applied to a simple
oscillation equation.  The analysis there shows that LT is more accurate
than SI for both the linear and nonlinear terms.
In \S\ref{sec:filter}, the filtering procedure is outlined.
In \S\ref{sec:scheme}, the application of the LT scheme to the {\Peak} model
is described in detail. It will be seen
that the scheme is of complexity comparable to the SI scheme. The
one additional process is a transformation between model levels and
vertical eigenmodes.

Some numerical experiments are described in \S\ref{sec:numerics}.
These include integrations
starting from several initial conditions: a {Kelvin wave}; the five-day wave;
a Rossby-Haurwitz wave RH(4,5); a perturbed unstable zonal jet leading to a
growing baroclinic wave; analysed meteorological data for a specific date and time.
In all cases, the LT scheme is of accuracy equal to or greater than the 
widely-used SI scheme. In particular, LT yields a more accurate 48 hour forecast
from real meteorological data.
Discussion of further advantages and potential of the LT scheme and conclusions 
may be found in \S\ref{sec:conclusions}.


\section{Accuracy Analysis}
\label{sec:anal}

We examine the properties of the SI and LT schemes by applying them
to a simple nonlinear oscillation equation
\begin{equation}
\frac{\dd X}{\dd t} = i\omega X + N(X)
\label{eq:oscilleqn}
\end{equation}
We assume that the nonlinear term consists of high-frequency variations around a
steady value, $N(t) = \bar N + N^\prime(t)$ and,
for the present analysis, we approximate $N$ by $\bar N$. The exact solution is then
$$
X(t) = X^{0}\exp(i\omega t) + \left[\frac{\exp(i\omega t)-1}{i\omega}\right]{\bar N}
$$
We express the solution $X^{+}$ at time $(n+1)\Delta t$ as
\begin{equation}
X^{+} = \biggl[\exp(2i\theta)\biggr] X^{-} 
 + \left[\frac{\exp(2i\theta)-1}{2i\theta}\right]{2\Delta t \bar N}
\label{eq:exactsoln}
\end{equation}
where $X^{-}$ is the solution at time $(n-1)\Delta t$ and
we have introduced the digital frequency $\theta = \omega\Delta t$.

The values produced by the SI and LT schemes will be compared to the exact solution.
The SI appproximation for (\ref{eq:oscilleqn}) is
\begin{equation}
\frac{X^{+}-X^{-}}{2\Delta t} =
i\omega\frac{X^{+}+X^{-}}{2} + \bar N
\nonumber
\end{equation}
Solving for the new value $X^{+}$, we have
\begin{equation}
X^{+} = \left(\frac{1+i\theta}{1-i\theta}\right)X^{-}
        + \left(\frac{1}{1-i\theta}\right){2\Delta t \bar N}
\label{eq:SIaxe}
\end{equation}
Comparing this with the exact solution (\ref{eq:exactsoln}), we see that both the linear 
and nonlinear components of the solution are misrepresented.
For the exact solution (\ref{eq:exactsoln}), $X^{-}$ is multiplied by $\exp(2i\theta)$,
which has unit modulus and phase $2\theta$. For the SI solution (\ref{eq:SIaxe}),
the multiplier is
$$
\rho = \left[\frac{1+i\theta}{1-i\theta}\right] \,,
$$
which has modulus and phase given by
$$
|\rho| = 1 \qquad\mbox{and}\qquad \arg\rho = 2\arctan\theta \,.
$$
Thus, while there is no amplification, there is a phase error 
that increases with the value of the digital frequency $\theta$.

The multiplier of the nonlinear term $2\Delta t \bar N$ in (\ref{eq:exactsoln}) is
$$
\rho = \left[\frac{\exp(2i\theta)-1}{2i\theta}\right] \,,
$$
with modulus and phase of $\rho$ given by
$$
|\rho| = \frac{\sin \theta}{\theta} \qquad\mbox{and}\qquad \arg\rho = \theta
$$
The corresponding factor for the SI scheme (\ref{eq:SIaxe}) is
$$
\rho = \left[\frac{1}{1-i\theta}\right] \,,
$$
with modulus and phase of $\rho$ given by
$$
|\rho| = \frac{1}{\sqrt{1+\theta^2}} \qquad\mbox{and}\qquad \arg\rho = \arctan\theta \,.
$$
Thus, the SI scheme has both modulus and phase errors in the treatment of the nonlinear term.

We now apply the Laplace Transform $\cal L$ to the equation (\ref{eq:oscilleqn}),
taking the origin of time to be $(n-1)\Delta t$:
\[
s\hat{X} - X^{-} = i\omega\hat{X} + \frac{\bar N}{s}.
\]
Solving for $\hat{X}$ gives
\begin{equation}
\hat{X} = \frac{X^{-}}{s-i\omega} + \frac{\bar N}{s(s-i\omega)} \,.
\label{eq:XplusLT}
\end{equation}
The inverse Laplace transform ${\cal L}^{-1}$ with time set to $2\Delta t$ yields
the exact solution (\ref{eq:exactsoln}) at time $(n+1)\Delta t$.

\section{Filtering with the LT Scheme}
\label{sec:filter}

The LT scheme filters high frequency components by using a modified inversion
operator ${\cal L}^{*}$: this can be done numerically by distorting the Bromwich countour for the
inversion integral to a closed curve excluding poles associated with the high frequencies,
as in Clancy and Lynch (2011a, b). 

In this paper, as in Lynch and Clancy (2016), we invert the
transform analytically, explicitly eliminating components with frequency greater than a specified
cut-off frequency $\omega_c$. This may be done with a sharp cut-off at $\omega_c$, or with
a smooth function such as a Butterworth filter having frequency response function
\begin{equation}
{\cal H}(\omega) = \frac{1}{1+(\omega/\omega_c)^L}
\label{eq:butter}
\end{equation}
Formally, we can define the modified inversion operator as the composition of
the filter and the inverse Laplace transform: ${\cal L}^{*} = {\cal L}^{-1}{\cal H}$.

In the numerical tests reported in \S\ref{sec:numerics} we set $L=16$.
We are free to choose the cut-off frequency. In the numerical experiments, we
chose a cut-off period $\tau_c=1\,$h and set $\omega_c=2\pi/\tau_c$.
Assuming that $|\omega|\ll\omega_c$, the inverse Laplace transform
of $(\ref{eq:XplusLT})$ at time $(n+1)\Delta t$ gives
\[
X^{+} = \left[ \exp(2i\omega\Delta t) \right] X^{-}
      + \left[\frac{\exp(2i\omega\Delta t) - 1}{i\omega}\right] \bar N\,.
\]
This agrees with the exact analytic result (\ref{eq:exactsoln}) for both
the linear and nonlinear terms. Thus, the LT scheme is free from error
(to the extent that the nonlinear term can be regarded as constant).



\section{Laplace transforming the {\Peak} Equations}
\label{sec:scheme}

Our objective is to implement a Laplace transform time integration
scheme in the {\Peak} Model. We begin with the equations (9.38)--(9.41) 
in \S9.3 of Ehrendorfer (2012):
\begin{eqnarray}
\frac{\partial\bfeta}{\partial t} &=& {\bff}_{\bfeta} - \varsigma\bfeta
\label{eq:voreqn}
\\
\frac{\partial\bfdelta}{\partial t} &=& {\bff}_{\bfdelta} 
-\bfDelsq(\bfTbar\pi + \bfPhi_{*} + \ssG\bfT) - \varsigma\bfdelta 
\label{eq:deleqn}
\\
\frac{\partial\bfT}{\partial t} &=& {\bff}_{\bfT} - \ssH\bfdelta - \varsigma\bfT
\label{eq:Teqn}
\\
\frac{\partial\pi}{\partial t} &=& {f}_{\pi} - \bfpT\bfdelta
\label{eq:pieqn}
\end{eqnarray}
The notation is generally conventional and equations similar to these have appeared
frequently, going back to Hoskins and Simmons (1975).
The dependent variables are vorticity ($\bfeta$),
divergence ($\bfdelta$), temperature ($\bfT$) and log surface pressure
$\pi=\log(p_s/p_{*}$), where $p_{*}=10^5\,$Pa is the reference pressure.
All variables are in the spectral domain but, for compactness, we suppress the spectral indices
so that, for example, ${\bfeta}^m_\ell$ is written simply as ${\bfeta}$. Bold-face variables 
are vectors with values at all model levels. The explicit
expressions for the matrices $\ssG$ and $\ssH$ and vectors $\bfTbar$ and $\bfpT$ are given by
Ehrendorfer (2012). Linear damping with coefficient $\varsigma$ is applied to all variables
except the surface pressure.

If nonlinear terms, orography and damping are omitted, the vorticity equation is
${\bfeta}_{t} = 0$ and vorticity is linearly decoupled from the other variables.
The remaining three equations are
\begin{equation}
\frac{\partial}{\partial t}
\begin{pmatrix} \bfdelta \\ \bfT \\ \pi \end{pmatrix}
=
\begin{bmatrix}
   \bfzero  &  -\bfDelsq\ssG & -\bfDelsq\bfTbar \\
    - \ssH  &  \bfzero             & \bfzero \\
    - \bfpT &  \bfzero             &    0    \\
\end{bmatrix}
\begin{pmatrix} \bfdelta \\ \bfT \\ \pi \end{pmatrix}
\end{equation}

We now take the Laplace transform of (\ref{eq:deleqn}), (\ref{eq:Teqn}) and (\ref{eq:pieqn}),
assuming centered approximations for the nonlinear terms and a forward step for diffusion:
\begin{eqnarray}
s\hat\bfdelta &=& {\bff}_{\bfdelta}/s
-\bfDelsq(\bfTbar\hat\pi + \bfPhi_{*}/s + \ssG\hat\bfT) + (1-\varsigma/s)\bfdelta^{-}   
\nonumber   
\\
s\hat\bfT &=& {\bff}_{\bfT}/s - \ssH\hat\bfdelta + (1-\varsigma/s)\bfT^{-}
\label{eq:eTem1}
\\
s\hat\pi &=& {f}_{\pi}/s - \bfpT\hat\bfdelta + \pi^{-}
\label{eq:epi1}
\end{eqnarray}
Using the equations for $\hat\bfT$ and $\hat\pi$, these quantities are
eliminated from the divergence equation to obtain an equation
for a single variable, $\hat\bfdelta$:
\begin{eqnarray*}
s^2\hat\bfdelta
&=& {\bff}_{\bfdelta} -\bfDelsq(\bfTbar s\hat\pi
+ \bfPhi_{*} + \ssG s\hat\bfT) + (s-\varsigma)\bfdelta^{-}
\\
&=& {\bff}_{\bfdelta} - \bfDelsq[\bfPhi_{*}]
- \bfDelsq[\bfTbar({f}_{\pi}/s - \bfpT\hat\bfdelta + \pi^{-})
\\ &  &
- \bfDelsq[\ssG ({\bff}_{\bfT}/s - \ssH\hat\bfdelta + (1-\varsigma/s)\bfT^{-})]
\\ &  &
+ (s-\varsigma)\bfdelta^{-}
\end{eqnarray*}
We transfer all terms involving $\hat\bfdelta$ to the left:
\begin{eqnarray}
[s^2\ssI &-& \bfDelsq(\bfTbar\bfpT + \ssG\ssH)]\hat\bfdelta =
\label{eq:delhat1}
\\ &  & 
{\bff}_{\bfdelta} - \bfDelsq[\bfPhi_{*}]
- (1/s)\bfDelsq[\bfTbar{f}_{\pi} + \ssG{\bff}_{\bfT}]
\nonumber
\\ &  &
- \bfDelsq[\bfTbar\pi^{-} + (1-\varsigma/s)\ssG\bfT^{-}]
+ (s-\varsigma)\bfdelta^{-}
\nonumber
\end{eqnarray}

\subsection*{Transforming to Vertical Eigenmodes}

We now define the vertical structure matrix
$$
\ssB = \bfTbar\bfpT + \ssG\ssH \,.
$$
Since $\ssB$ is symmetric, its eigen-structure can be written
$$
\ssB\ssE = \ssE\ssLAM \qquad\mbox{or}\qquad
\ssET\ssB\ssE = \ssLAM \qquad\mbox{or}\qquad
\ssB = \ssE\ssLAM\ssET
$$
We now transform to vertical eigenmodes by multiplying (\ref{eq:delhat1}) by $\ssET$:
\begin{eqnarray}
&& \bigl[s^2\ssI - \bfDelsq(\ssET\ssB\ssE)\bigr](\ssET\hat\bfdelta) =
\label{eq:delhat2}
\\ && 
\quad \ssET \biggl\{ {\bff}_{\bfdelta} - \bfDelsq\bfPhi_{*} 
- \bfDelsq[\bfTbar{f}_{\pi} + \ssG{\bff}_{\bfT}] (1/s)
\nonumber
\\ &&
\quad\qquad + \bfdelta^{-}(s-\varsigma) 
- \bfDelsq[\bfTbar\pi^{-} + (1-\varsigma/s)\ssG\bfT^{-}] 
\biggr\}
\nonumber
\end{eqnarray}
We note that, for a specific spectral component with total wavenumber
$n$, the laplacian has a simple form $ - \bfDelsq = n(n+1)/a^2$.
Then, defining $ - \bfDelsq\ssLAM = \ssOM^2$,
the matrix on the left hand side of (\ref{eq:delhat2}) is
$$
[s^2\ssI - \bfDelsq(\ssET\ssB\ssE)] =
[s^2\ssI - \bfDelsq\ssLAM] =
[s^2\ssI + \ssOM^2]
$$
Since this is a diagonal matrix, the equation falls into ${\cal M}$ separate
scalar equations, one for each vertical mode.  We group the right hand
terms of (\ref{eq:delhat2}) according to powers of $s$:
\begin{equation}
\mbox{RHS} = 
\ssET\bigl\{
 {\cal A}\times s + {\cal B}\times 1 + {\cal C}\times (1/s) \bigr\}
\label{eq:ABCeqn}
\end{equation}
where the vectors $\cal A$, $\cal B$ and $\cal C$ are
\begin{eqnarray*}
{\cal A} &=&  {\bfdelta}^{-} 
\\
{\cal B} &=& 
 {\bff}_{\bfdelta} - \bfDelsq\bfPhi_{*} - \varsigma{\bfdelta}^{-}
- \bfDelsq(\bfTbar\pi^{-} + \ssG\bfT^{-})
\\
{\cal C} &=& 
  - \bfDelsq(\bfTbar{f}_{\pi} + \ssG{\bff}_{\bfT} - \varsigma\ssG\bfT^{-})
\end{eqnarray*}
Multiplying (\ref{eq:delhat2}) by the inverse of the
diagonal matrix $s^2\ssI + \ssOM^2$, the equation for the $k$-th component is
\begin{eqnarray}
(\ssET\hat\bfdelta)_k &=&
\left[ \frac{s}{s^2+\Omega_k^2} \right] (\ssET{\cal A})_k 
\nonumber
\\
&+& \left[ \frac{1}{s^2+\Omega_k^2} \right] (\ssET{\cal B})_k 
\nonumber
\\
&+& \left[ \frac{1}{s(s^2+\Omega_k^2)} \right] (\ssET{\cal C})_k 
\label{eq:delhatbird}
\end{eqnarray}
The terms of (\ref{eq:delhatbird}) can be inverted by means of
the following standard results:
\begin{eqnarray}
{\cal L^{*}}\left\{\frac{s}{s^2+\Omega_k^2} \right\} &=&
{\cal H}(\Omega_k)\cos\Omega_k t
\label{eq:inva}
\\
{\cal L^{*}}\left\{\frac{1}{s^2+\Omega_k^2} \right\} &=&
\frac{{\cal H}(\Omega_k)\sin\Omega_k t}{\Omega_k}
\label{eq:invb}
\\
{\cal L^{*}}\left\{ \frac{1}{s(s^2+\Omega_k^2)} \right\} &=&
\frac{1 - {\cal H}(\Omega_k)\cos\Omega_k t}{\Omega_k^2}
\label{eq:invc}
\\
{\cal L^{*}}\left\{ \frac{1}{s^2(s^2+\Omega_k^2)} \right\} &=&
\frac{\Omega_k t - {\cal H}(\Omega_k)\sin\Omega_k t}{\Omega_k^3}
\label{eq:invd}
\end{eqnarray}
((\ref{eq:invd}) will be used below).
The response function ${\cal H}(\omega)$ was defined in equation (\ref{eq:butter}).

We apply the operator ${\cal L^{*}}$ to (\ref{eq:delhatbird}),
noting that the vertical transform and Laplace transform commute.
The result at time $2\Delta t$ is denoted by a $+$-superscript:
\begin{eqnarray}
(\ssET\bfdelta)_k^{+} &=&  
\left[ {\cal H}(\Omega_k)\cos 2\Omega_k\Delta t \right] (\ssET{\cal A})_k 
\label{eq:delbird}
\\ &+& 
\left[ \frac{{\cal H}(\Omega_k)\sin 2\Omega_k\Delta t}{\Omega_k} \right] (\ssET{\cal B})_k
\nonumber
\\ &+& 
\left[ \frac{1-{\cal H}(\Omega_k)\cos 2\Omega_k\Delta t}{\Omega_k^2} \right] (\ssET{\cal C})_k 
\nonumber
\end{eqnarray}

\subsection*{Inverse Vertical Transformation}

Let us define four diagonal matrices
\begin{eqnarray*}
\ssLAM_{A} &=& \diag\left({\cal H}(\Omega_k)\cos 2\Omega_k\Delta t \right)
\\
\ssLAM_{B} &=& \diag\left( \frac{{\cal H}(\Omega_k)\sin 2\Omega_k\Delta t}{\Omega_k} \right)
\\
\ssLAM_{C} &=& \diag\left(
\frac{1-{\cal H}(\Omega_k)\cos 2\Omega_k\Delta t}{\Omega_k^2} \right)
\\
\ssLAM_{D} &=& \diag\left(
\frac{2\Omega_k\Delta t-{\cal H}(\Omega_k)\sin 2\Omega_k\Delta t}{\Omega_k^3} \right)
\end{eqnarray*}
($\ssLAM_{D}$ will be needed below). Then (\ref{eq:delbird}) can be written
\begin{equation}
(\ssET\bfdelta)^{+} = 
\ssLAM_A\ssET{\cal A} + \ssLAM_B\ssET{\cal B} + \ssLAM_C\ssET{\cal C}
\label{eq:delbrd2}
\end{equation}
We can now calculate the divergence at the advanced time,
\begin{eqnarray*}
\bfdelta^{+} &=& \ssE(\ssET\bfdelta)^{+} 
\\ &=&
\ssE\ssLAM_A\ssET{\cal A} + \ssE\ssLAM_B\ssET{\cal B} + \ssE\ssLAM_C\ssET{\cal C}
\end{eqnarray*}
For compactness, we define the propagation matrices:
\begin{eqnarray*}
\ssPA =  \ssE\ssLAM_A\ssET &\qquad& \ssPB =  \ssE\ssLAM_B\ssET \\
\ssPC =  \ssE\ssLAM_C\ssET &\qquad& \ssPD =  \ssE\ssLAM_D\ssET
\end{eqnarray*}
($\ssPD$ will be used below). Then we can write the solution as 
\begin{equation}
\bfdelta^{+} = \ssPA{\cal A} + \ssPB{\cal B} + \ssPC{\cal C}
\label{eq:deltaplus}
\end{equation}
The $\ssP$-matrices can be pre-computed and stored, since they
do not depend on the model variables.

\subsection*{Temperature and Pressure: Method I}

We describe two methods of computing the temperature and pressure fields
at time $(n+1)\Delta t$ using $\bfdelta^{+}$. 
In Method I, we use (\ref{eq:eTem1}) and (\ref{eq:epi1}) to compute $\bfT^{+}$ and $\pi^{+}$.
Dividing (\ref{eq:eTem1}) by $s$ and applying the operator ${\cal L}^{*}$
at time $2\Delta t$, we have
$$
\bfT^{+} = \bfT^{-} + 
2\Delta t( {\bff}_{\bfT} - \varsigma\bfT^{-}) 
- \ssH{\cal L}^{*}\left\{\hat\bfdelta/s \right\}
$$
The final term inverts to an integral that is approximated
by the trapezoidal rule
$$
{\cal L}^{*}\left\{\hat\bfdelta/s \right\}
= \int_0^{2\Delta t} \delta \, \dd t
\approx 2\Delta t\left(\frac{\bfdelta^{-}+\bfdelta^{+}}{2}\right)
= 2\Delta t\bfdelbar
$$
where $\bfdelbar$ is the average of old and new values of $\bfdelta$.
Then the temperature at the new time is
\begin{equation}
\bfT^{+} = \bfT^{-} 
           + 2\Delta t ( {\bff}_{\bfT} -\varsigma\bfT^{-} - \ssH\bfdelbar )
\label{eq:Tplus}
\end{equation}
In a similar way, (\ref{eq:epi1}) transforms to give
\begin{equation}
\pi^{+} = \pi^{-} + 2\Delta t ( {f}_{\pi} - \bfpT\bfdelbar )
\label{eq:piplus}
\end{equation}
Now $\bfdelta$, $\bfT$ and $\pi$ are available at the new time.

\subsection*{Temperature and Pressure: Method II}

We return to equations (\ref{eq:eTem1}) and (\ref{eq:epi1}):
\begin{eqnarray*}
s\hat\bfT &=& {\bff}_{\bfT}/s - \ssH\hat\bfdelta + (1-\varsigma/s)\bfT^{-}
\nonumber
\\
s\hat\pi &=& {f}_{\pi}/s - \bfpT\hat\bfdelta + \pi^{-}
\end{eqnarray*}
Dividing (\ref{eq:eTem1}) and (\ref{eq:epi1}) by $s$ and applying the operator ${\cal L}^{*}$
at time $2\Delta t$, we have
\begin{eqnarray}
\bfT^{+} &=& \bfT^{-} + 
2\Delta t({\bff}_{\bfT} - \varsigma\bfT^{-}) 
- \ssH{\cal L}^{*}\left\{\hat\bfdelta/s \right\}
\label{eq:eTem2}\\
\pi^{+} &=& \pi^{-} \,+ 2\Delta t {f}_{\pi}- \bfpT{\cal L}^{*}\left\{\hat\bfdelta/s \right\}
\label{eq:epi2}.
\end{eqnarray}
Both (\ref{eq:eTem2}) and (\ref{eq:epi2}) require computation of 
\begin{equation}
\tilde{\bfdelta} = {\cal L}^{*}\left\{\hat\bfdelta/s \right\}.
\nonumber  
\end{equation}
This can be computed by noting that
\begin{equation}
\tilde{\bfdelta} = {\cal L}^{*}\left\{\hat\bfdelta/s \right\} 
                 =  \ssE{\cal L}^{*}\left\{\ssET\hat\bfdelta/s \right\}\,.
\label{eq:deltilde}
\end{equation}
We divide (\ref{eq:delhatbird}) by $s$ to give
\begin{eqnarray}
\left(\frac{\ssET\hat\bfdelta}{s}\right)_k
&=&  
\left[ \frac{1}{s^2+\Omega_k^2} \right] (\ssET{\cal A})_k 
\nonumber \\
&+&
\left[ \frac{1}{s(s^2+\Omega_k^2)} \right] (\ssET{\cal B})_k 
\nonumber \\
&+&
\left[ \frac{1}{s^2(s^2+\Omega_k^2)} \right] (\ssET{\cal C})_k 
\label{eq:tildel2}
\end{eqnarray}
We invert this using the standard results (\ref{eq:invb})--(\ref{eq:invd}) to give
\begin{eqnarray*}
{\cal L}^{*}\left\{\frac{\ssET\hat\bfdelta}{s} \right\}_k &=&
\left[ \frac{{\cal H}(\Omega_k)\sin 2\Omega_k\Delta t}{\Omega_k} \right] (\ssET{\cal A})_k 
\\
+&& \hspace{-5mm}
\left[ \frac{1-{\cal H}(\Omega_k)\cos 2\Omega_k\Delta t}{\Omega_k^2} \right]
(\ssET{\cal B})_k 
\\
+&& \hspace{-5mm}
\left[ \frac{2\Omega_k\Delta t-{\cal H}(\Omega_k)\sin 2\Omega_k\Delta t}{\Omega_k^3} \right]
(\ssET{\cal C})_k 
\end{eqnarray*}
Then using the $\Lambda$-matrices, we can write
\begin{equation}
{\cal L}^{*}\left\{\frac{\ssET\hat\bfdelta}{s} \right\} 
= \Lambda_B\ssET{\cal A} + \Lambda_C\ssET{\cal B} + \Lambda_D\ssET{\cal C}
\nonumber
\end{equation}
Noting (\ref{eq:deltilde}) and using the $P$-matrices, we can now write
\begin{equation}
\tilde{\bfdelta} =  \ssPB{\cal A} + \ssPC{\cal B} + \ssPD{\cal C} \,.
\end{equation}
Finally, using (\ref{eq:eTem2}) and (\ref{eq:epi2}), the values of $T^{+}$ and $\pi^+$ are
\begin{eqnarray}
\bfT^{+} &=& \bfT^{-} + 
2\Delta t( {\bff}_{\bfT} - \varsigma\bfT^{-} -  \ssH\tilde{\bfdelta}/2\Delta t)
\label{eq:eTemplus}\\
\pi^{+} &=& \pi^{-} \,+ 2\Delta t ({f}_{\pi} - \bfpT\tilde{\bfdelta}/2\Delta t) \,.
\label{eq:epiplus}
\end{eqnarray}

\subsection*{Integrating the Vorticity}

To complete the timestep, we integrate the vorticity equation 
(\ref{eq:voreqn}) with a leapfrog timestep for the nonlinear terms and a
forward step for diffusion:
\begin{equation}
\bfeta^{+} = \bfeta^{-} + 2\Delta t ({\bff}_{\bfeta} - \varsigma\bfeta^{-} )
\label{eq:etaplus}
\end{equation}
We now have all the model variables at the advanced time, and a new timestep can be
taken.

As usual with the leapfrog model, a Robert-Asselin filter is applied to prevent
separation of the solutions at odd and even timesteps. The coefficient is 
fixed at $\epsilon = 0.03$ in all cases.


\section{Numerical Experiments}
\label{sec:numerics}

We describe here a series of tests of the Laplace transform (LT) scheme.
The reference runs use the semi-implicit (SI) scheme with a one-minute time step.
No explicit horizontal diffusion was used for the Kelvin Wave and Five-day Wave
solutions. For the baroclinic wave and the real data, the damping coefficient was
set at $\varsigma = 7\times 10^{5}$: the damping of a component of total wavenumber
$n$ is $\varsigma n(n+1)/a^2$. Damping was also found to be necessary to stabilize the SI
integration of the Rossby-Haurwitz Wave RH(4,5); no diffusion was necessary for the LT scheme.

The SI and LT models were run with 10 and 20 minute timesteps
and scored against the reference. For LT, the cut-off period of one hour
for $\Delta t=600$s may be scaled in proportion to the time step ($\tau_c=6\Delta t$).
However, we used a value $\tau_c=1\,$h for both timesteps (10 and 20 minutes).

Numerical tests indicated clearly that, of the two methods described in 
\S\ref{sec:scheme} for advancing the temperature and log surface pressure,
Method~II was clearly superior. Therefore, this method was used for all the
experiments described below.

For the planetary wave conditions, the horizontal resolution of the model is at
triangular truncation $T85$.  The colocation grid has $256\times 129$ grid points,
corresponding to a grid interval of approximately $150$km. For an advection speed $\bar u = 100$m/s, 
the nondimensional stability ratio $\bar u\Delta t / \delta x$ is unity for 
a time step $\Delta t = 1500$s or 25 minutes. In fact, both the SI and LT models
were found to be unstable for a time step of 24 minutes.

For the case of real atmospheric data, the spatial truncation was $T129$, and the
colocation grid had $512\times 259$ grid points.
In all cases, there were 20 vertical levels, uniformly spaced in $\sigma$-coordinates.


\subsection*{Kelvin Wave}

\begin{figure}[!htbp]
\centering
\includegraphics[width=0.75\textwidth]{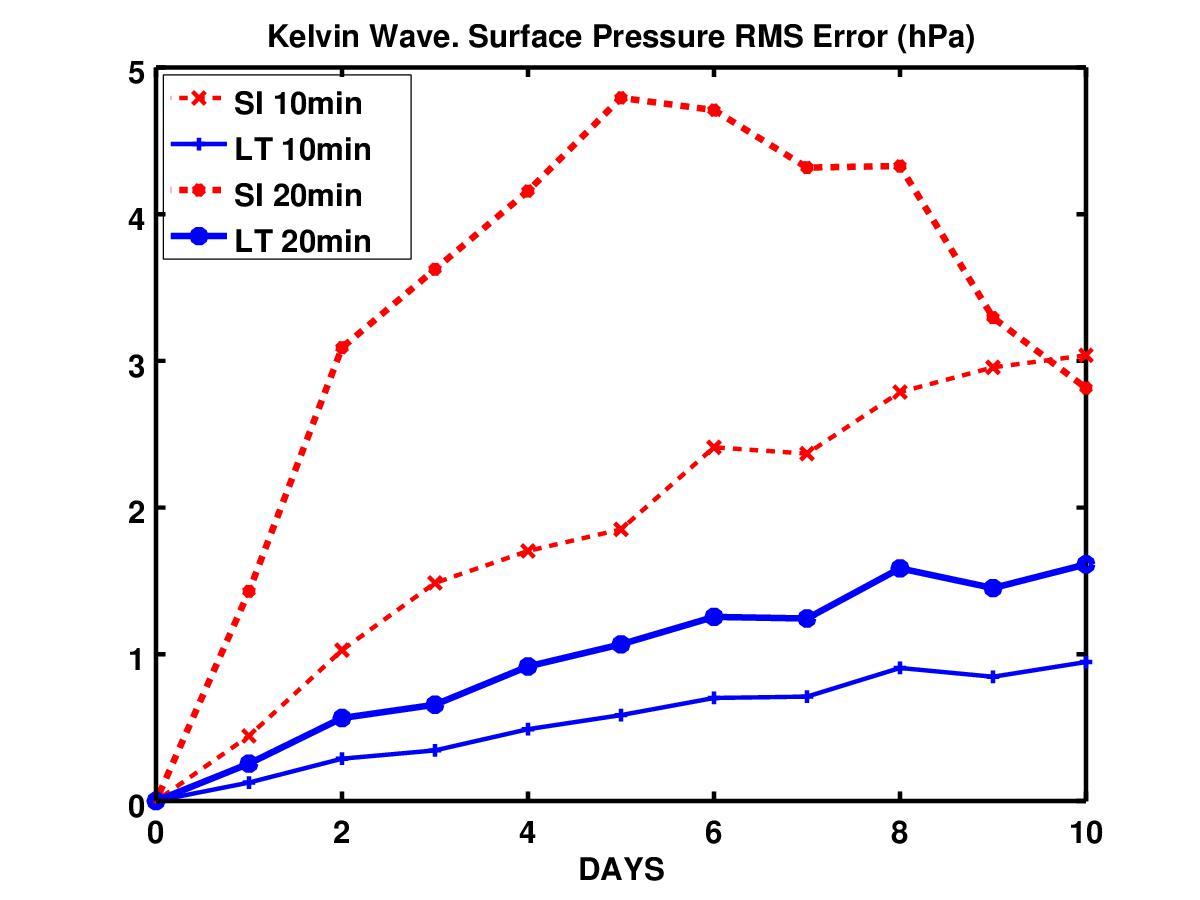}
\caption{Kelvin wave: rms error of surface pressure (hPa) for 10 day forecasts.
Solid lines: LT. Dashed lines: SI.
Thin lines: $\Delta t = 10\,$min. Thick lines: $\Delta t = 20\,$min.}
\label{fig:KELVIN}
\end{figure}

Kelvin waves are eastward propagating waves that play an important
role in atmospheric dynamics. Previous work (Clancy \&\
Lynch, 2011) showed that the LT scheme had a significantly smaller phase
error than the semi-implicit scheme for the integration of these waves.
Exact Kelvin Wave initial conditions can be generated using the method in
Kasahara (1976). In this study, we use a simple analytical approximation%
\footnote{Details in Harney, Eoghan, 2018:
\emph{Laplace Transform Integration Scheme in barotropic and baroclinic
atmospheric models}. PhD Thesis, University College Dublin.}.
The zonal wavenumber is $m=4$ and the wave
amplitude is 100\,m.  The theoretical period for the Kelvin wave with
zonal wavenumber 4 is about 8.3 hours (Kasahara, 1976, Fig.~9).
Fig.~\ref{fig:KELVIN} shows that the results for LT (solid lines)
are superior to SI (dashed lines). Forecasts were run with time steps of
10 minutes (thin lines) and 20 minutes (thick lines).
The phase error for the SI forecast with a 20 minute step is so large
that the propagation of the wave has lagged by almost a full wavelength
by the end of the integration.

\subsection*{The Five-Day Wave}

\begin{figure}[!htbp]
\centering
\includegraphics[width=0.75\textwidth]{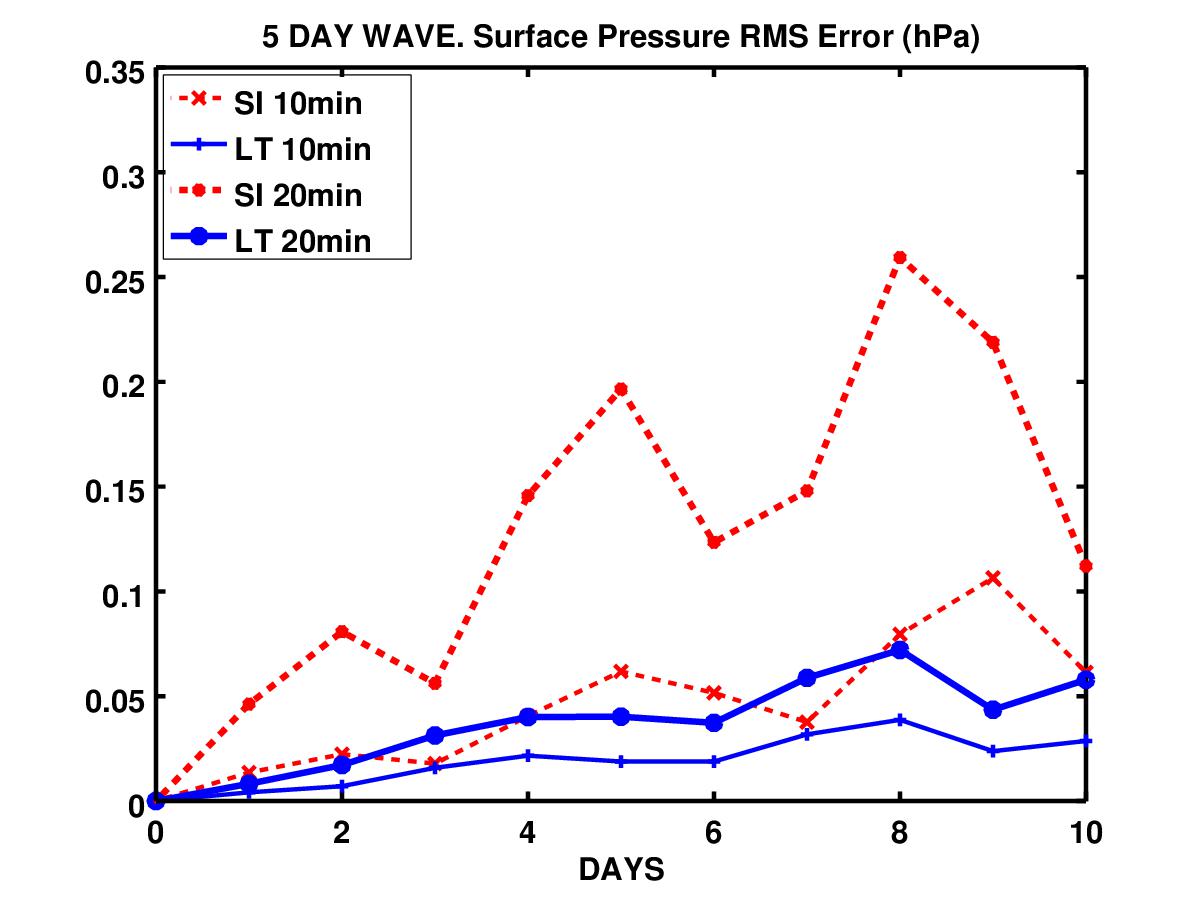}
\caption{Five-day wave RH(1,2): rms error of surface pressure (hPa) for 10 day forecasts.
         Solid lines: LT. Dashed lines: SI.  Thin lines: $\Delta t = 10\,$min.
         Thick lines: $\Delta t = 20\,$min.}
\label{fig:RH12-4PLOT-PS-rms}
\end{figure}

The Five Day Wave RH(1,2) is the gravest symmetric rotational Hough mode of zonal wavenumber 1.
It is close to the initial state chosen by Lewis Fry Richardson for his preliminary
shallow-water forecast experiment (Lynch, 2006).
The initial conditions for a three dimensional Five Day Wave were implemented in \Peak.
The initial pressure amplitude was 25hPa, with mean pressure 1000hPa.
As no zonal mean flow was included, the wave has a period close to 5 days.
Fig.~\ref{fig:RH12-4PLOT-PS-rms} shows the rms errors in surface pressure for the LT
scheme (solid) and the SI scheme (dashed), with timesteps of 10 minutes (thin lines) 
and 20 minutes (thick lines).
As usual, the reference is an SI forecast with time step 60s. 
The error level for LT is significantly less than that of the SI scheme,
especially for the longer timestep.
The pattern of scores for vorticity (not shown) is similar, 
indicating a substantial reduction in forecast error for LT compared to SI.

\subsection*{Rossby-Haurwitz Wave}

Rossby-Haurwitz (RH) waves are exact solutions of the nonlinear barotropic vorticity equation.
While they are not eigenfunctions of the shallow water equations, 
they have frequently been used as test cases. Following Phillips (1959), the RH(4,5) wave
was chosen as Test Case 6 by Williamson et al.~(1992). 
%
%
This test case has been extended to three dimensions;
the initial vorticity field is as in the barotropic case, the divergence is zero and
a vertical temperature profile and surface pressure field are defined;
for details, see Jablonowski et al.~(2008).
%
%
The wave was defined as a test case in the \Peak\ model.
The SI and LT schemes, with time steps of 10 and 20 minutes, were compared to a reference
run of SI using a time step of 60s. 
No diffusion was used for the reference or LT runs, but the SI runs were unstable.
This was overcome by applying horizontal damping with a coefficient
$\varsigma = 3\times10^{6}$.

\begin{figure}[!htbp]
\centering
\includegraphics[width=0.75\textwidth]{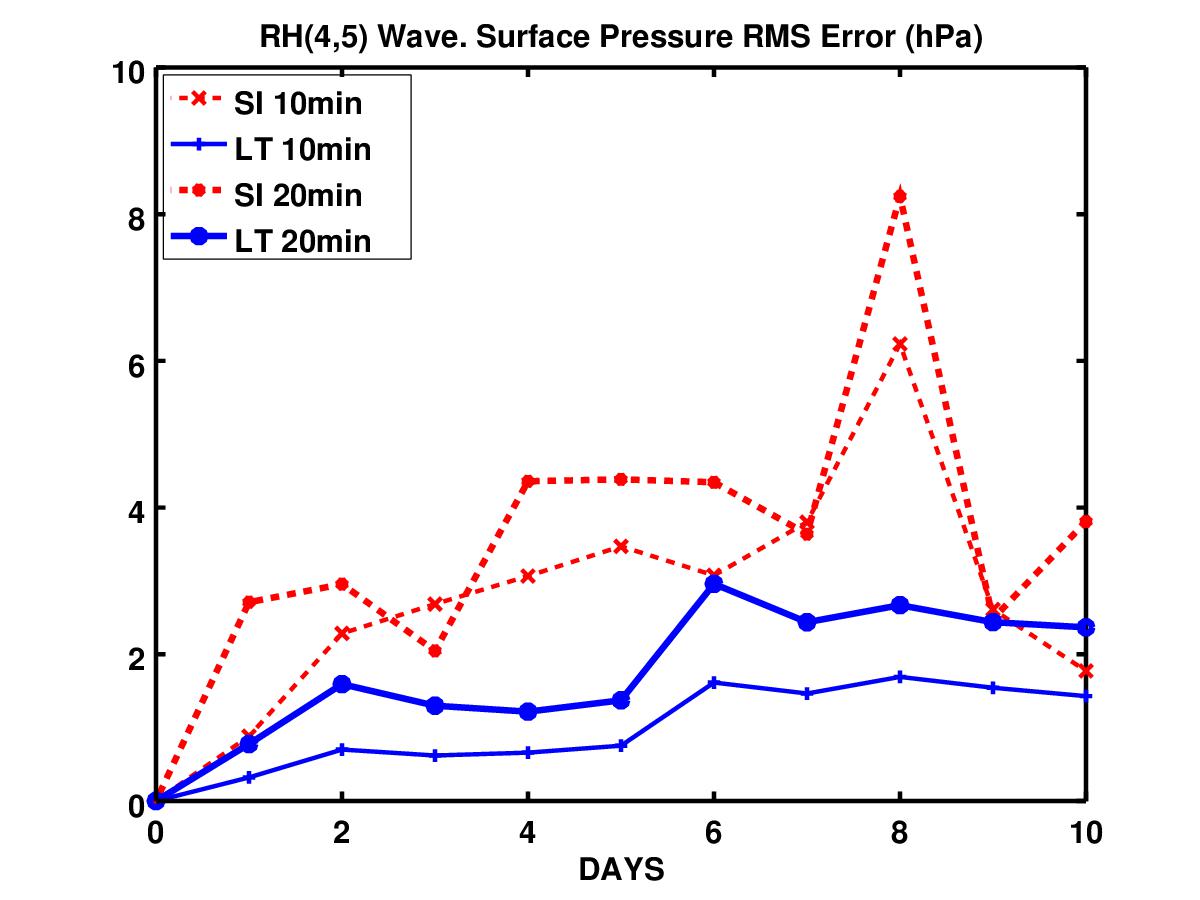}
\caption{RH(4,5): rms error of surface pressure (hPa) for 10 day forecasts.
         Solid lines: LT. Dashed lines: SI.  Thin lines: $\Delta t = 10\,$min.
         Thick lines: $\Delta t = 20\,$min.}
\label{fig:RH45-4PLOT-PS-rms}
\end{figure}

Fig.~\ref{fig:RH45-4PLOT-PS-rms} shows the
rms errors in surface pressure for the LT scheme (solid) and the SI scheme (dashed),
with timesteps of 10 minutes (thin lines) and 20 minutes (thick lines).
The error level for LT is substantially less than that of the SI scheme.
Scores for vorticity near the tropopause (model level 5, $\sigma = 0.225$,
not shown) are similar in pattern, confirming the superior accuracy of the LT scheme.

\subsection*{Baroclinic Development}

Polvani et al.~(2004) devised a test case for baroclinic instability.
The initial conditions consist of a nondivergent zonal flow with constant surface pressure.
The temperature profile is defined by using the meridional momentum equation with the
hydrostatic balance relation to give the meridional derivative of temperature.
This is numerically integrated to give the initial temperature field.
A small perturbation of the temperature is added to trigger the development of 
baroclinic instability.
With a fixed value for the diffusion coefficient, the initial conditions are
`numerically convergent' as shown by Polvani et al.~(2004) using two different
numerical models.  A test case quite similar to that of Polvani et~al.~was constructed
by Jablonowski and Williamson (2006).

\begin{figure}[!htbp]
\centering
\includegraphics[width=0.75\textwidth]{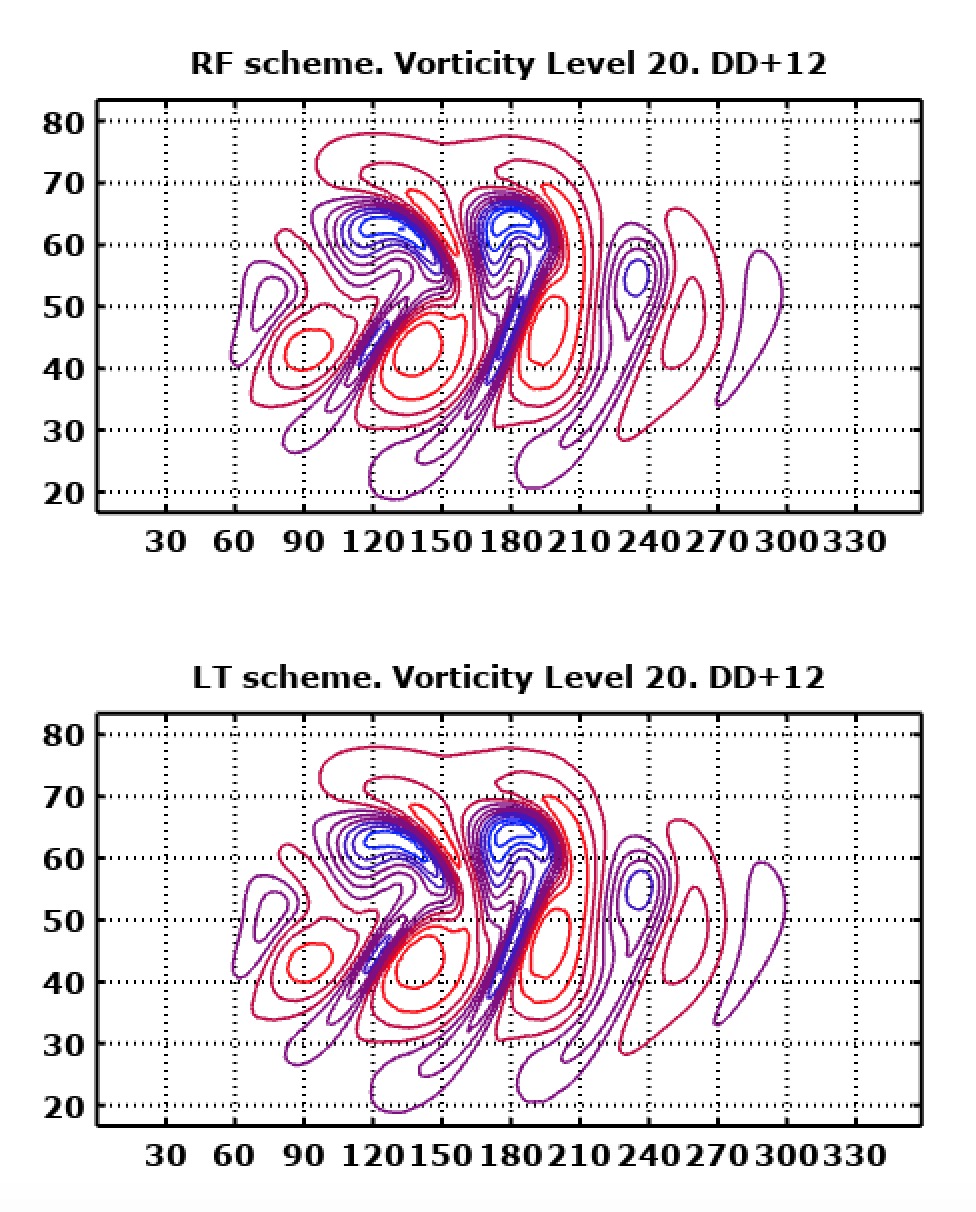}
\caption{Vorticity field at model level 20 for two 12-day forecasts.
Top panel. Reference, SI forecast with timestep 1 minute. 
Bottom panel. LT forecast with timestep 20 minutes.}
\label{fig:POL-RF-LT}
\end{figure}

The test case of Polvani et al.~was used by Ehrendorfer (2012) to validate the \Peak\ model.
We use it here to show that the LT scheme can accurately simulate baroclinic development.
Fig.~\ref{fig:POL-RF-LT} shows the vorticity field (at model level 20) at 10 days 
for a developing baroclinic wave. The reference forecast (top panel) uses the SI scheme
with a 1 minute timestep. The LT forecast (bottom panel) had a timestep of 20 minutes.
The differences between the two forecasts are negligible. They are also indistinguishable 
from the results plotted in Polvani et al.~(2004, Fig.~4). Thus, the LT scheme is capable
of forecasting baroclinic development with high precision. Quantitative scores confirm that the
error levels for LT and SI runs with $\Delta t=20\,$min are comparable and small.


\subsection*{Real Data and Initialization}

The simple wave tests described above indicate superior accuracy for the LT scheme
compared to the semi-implicit scheme.  The ultimate conclusion on superiority
of the scheme must involve comprehensive comparisons for a large range of meteorological
conditions. As a first step, a single test using real atmospheric data is described here.

Data was retrieved from the European Centre for Medium-Range Weather Forecasts MARS archive.
The date chosen was 00~UTC on 15th October, 2017, the day before storm Ophelia reached
Ireland. This data comprised temperature, divergence and vorticity fields on 25 pressure levels,
and surface pressure, at spectral triangular truncation of T129.
These fields where interpolated onto the 20 sigma levels of the \Peak\ model.
The process of interpolation introduced noise, 
which was removed by making a one-hour integration with the LT scheme to balance the
data (6 time steps of 600s, with $\tau_c=1$hr).

\begin{figure}[!h]
\centering
\includegraphics[width=0.75\textwidth]{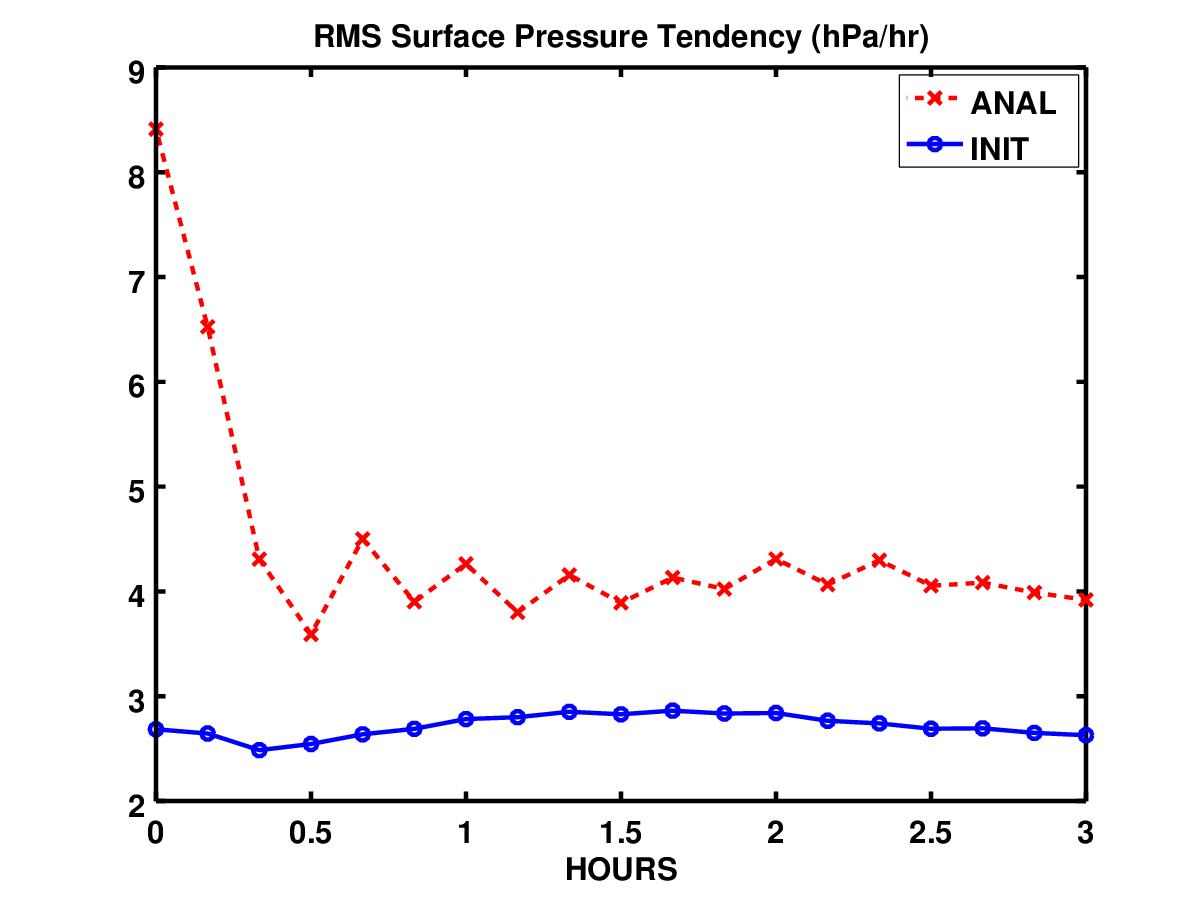}
\caption{$L_{2}$ norm of the tendency of $p_{s}$ (in hPa/hr) over 3 hours
         for uninitialized and initialized data. Both runs used the SI scheme
         with $\Delta t=600\,$s.}
\label{fig:Pi_Tend}
\end{figure}

Two 3-hour forecasts using the SI scheme were then made, one from the uninitialized
data and one from the initialized data. The level of gravity-wave noise is measured by
the $L_{2}$ norm of the tendency of surface pressure, $||{\partial p_{s}}/{\partial t}||_{2}$.
Fig.~\ref{fig:Pi_Tend} shows the tendency for an SI forecast from the analysis and
another from initialized data. Initialization effectively eliminates the noise that has
been introduced by interpolation.

\begin{figure}[!htbp]
\centering
\includegraphics[width=0.75\textwidth]{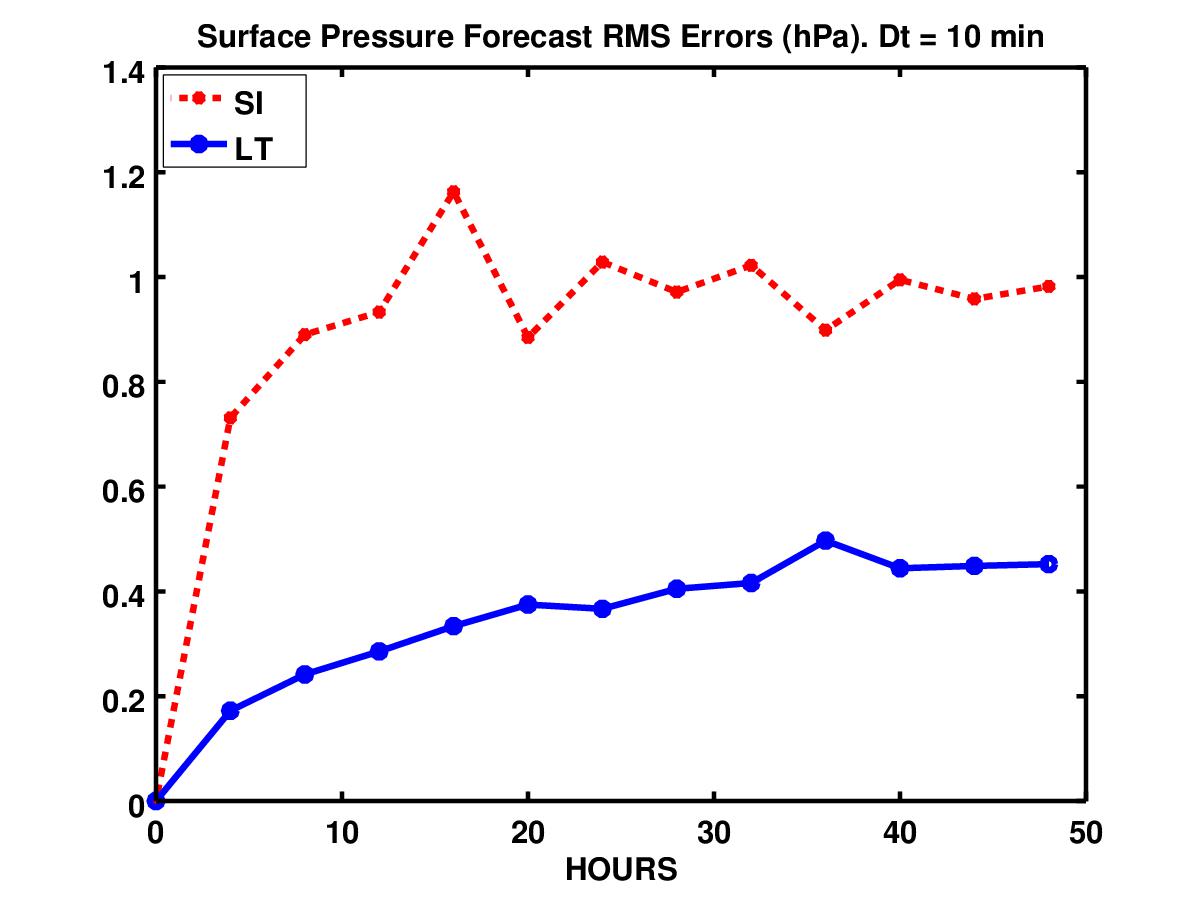}
\caption{Real data: maximum error for surface pressure (hPa) over 2 days.}
\label{fig:REAL-SILT-PS-max}
\end{figure}

Using initialized data, two forecasts were performed using SI and LT.
Fig.~\ref{fig:REAL-SILT-PS-max} shows the maximum error for surface pressure
for the SI and LT schemes, both with time step of 600s.
The reference is a forecast using SI with a time step of 60s.
The LT scheme produces a forecast of substantially greater accuracy than the SI scheme.
Scores for mid-troposphere vorticity (not shown) confirm this advantage.

\section{Concluding Remarks}
\label{sec:conclusions}

\subsection*{Future Work}

In preliminary work for this study, we combined the LT adjustment with a
semi-Lagrangian scheme for advection, using a shallow-water version of
OpenIFS. A Laplace/Lagrange scheme would have the potential to enable
larger time steps to be used while maintaining satsifactory forecast
accuracy. In the Lagrangian approach, the Laplace transform is calculated along
the time-dependent trajectory of a parcel of air. This means that,
contrary to a statement in Clancy and Lynch (2011b), we cannot assume
that ${\cal L}\nabla^2\Phi = \nabla^2{\cal L}\Phi$. The order of the
Laplace transform and spatial operators like the gradient cannot be
interchanged, and commutators such as $[{\cal L},\nabla^2]$ must be
taken into account; this remains to be done.

The LT method developed in \S\ref{sec:scheme} was based on a leapfrog
time-stepping scheme.  There are clear advantages associated with two time-level
schemes and it does not appear that there should be any difficulties
implementing an LT scheme in this context.

The limited numerical experimentation reported in \S\ref{sec:numerics}
indicated greater accuracy of LT compard to an SI scheme with the same
time step. In particular, the single real-data test forecast with LT was
more accurate. However, more extensive testing with a broad range of
initial meteorological states is required before definitive conclusions
on relative accuracy can be drawn.

\subsection*{Conclusion}

The LT scheme provides an attractive alternative to the popular
semi-implicit (SI) scheme.  The algorithmic complexity of the LT scheme
is comparable to SI, with just an additional transformation to vertical
eigenmodes each time step, and it provides the possibility of improving
weather forecasts at comparable computational cost.

\section{Acknowledgements}
\label{sec:acknowledgements}

We thank Glenn Carver of ECMWF for assistance in using the OpenIFS model at
a preliminary stage of this work. We are grateful to Martin Ehrendorfer for making 
the code of the \Peak\ model openly available.  EH acknowledges the financial
support of the Irish Research Council \\
(Grant No.~GOIPG/2015/2785).


\section*{References}
\begin{itemize}

\item[] 
Carver, Glenn, et al., 2018:
\emph{The ECMWF OpenIFS numerical weather prediction model release cycle 40r1:
description and use cases}. Manuscript in preparation.

\item[] 
Clancy, Colm and Peter Lynch, 2011a:
Laplace transform integration of the shallow water equations.
Part 1: Eulerian formulation and Kelvin waves.
\emph{Q.~J.~Roy.~Met.~Soc.}, {\bf 137}, 792--799.

\item[] 
Clancy, Colm and Peter Lynch, 2011b:
Laplace transform integration of the shallow water equations.
Part 2: Lagrangian formulation and orographic resonance.
\emph{Q.~J.~Roy.~Met.~Soc.}, {\bf 137}, 800--809.

%

\item[] 
Ehrendorfer, Martin, 2012: 
\emph{Spectral Numerical Weather Prediction Models.}
Soc.~Ind.~Appl.~Math. (SIAM), xxv+482 pp. ISBN: 978-1-61197-198-9



\item[] 
Hoskins, B.~J.~and A.~J.~Simmons, 1975:
A multi-layer spectral model and the semi-implicit method.
\emph{Q.~J.~Roy.~Met.~Soc.}, {\bf 101}, 637--655.

\item[] 
Jablonowski, C., P.~Lauritzen, R.~Nair and M.~Taylor, 2008:
Idealized test cases for the dynamical cores of Atmospheric General Circulation Models:
A proposal for the NCAR ASP 2008 summer colloquium.
{\tt http://www-personal.umich.edu/$\sim$cjablono/} \\
{\tt DCMIP-2008\underline{\ }TestCaseDocument\underline{\ }29May2008.pdf}

\item[] 
Jablonowski, C.~and D.~Williamson, 2006:
A baroclinic instability testcase for atmospheric model dynamical cores.
\emph{Q.~J.~Roy.~Met.~Soc.}, {\bf 132}, 2943--2975.

\item[] 
Kasahara A., 1976: Normal Modes of Ultralong Waves in the Atmosphere.
{\it Mon. Weather Rev.} {\bf 104}: 669--690.

\item[] 
Lynch, Peter, 1985a:
Initialization using Laplace transforms. \\
{\it Q.~J.~Roy.~Met.~Soc.}, {\bf 111}, 243--258.

\item[] 
Lynch, Peter, 1985b:
Initialization of a barotropic limited area model using the Laplace Transform technique.
{\it Mon.~Weather Rev.}, {\bf 113}, 1338--1344.


\item[] 
Lynch Peter, 2006:
\emph{The Emergence of Numerical Weather Prediction: Richardson's Dream}.
Cambridge University Press.

\item[] 
Lynch, Peter and Colm Clancy, 2016:
Improving the Laplace transform integration method
\emph{Q.~J.~Roy.~Met.~Soc.}, {\bf 142},1196-–1200. 

\item[] 
Phillips, Norman A., 1959:
Numerical integration of the primitive equations on the hemisphere.
\emph{Mon.~Wea.~Rev.}, {\bf 87}, 333--345.

\item[] 
Polvani, L,~M., R.~K.~Scott and S.~J.~Thomas, 2004:
Numerically Converged Solutions of the Global Primitive Equations for Testing the
Dynamical Core of Atmospheric GCMs.
\emph{Mon.~Wea.~Rev.}, {\bf 132}, 2539--2552.


\item[] 
Van Isacker J.~and W.~Struylaert, 1986:
Laplace Transform Applied to a Baroclinic Model. In {\it Short- and Medium-Range
Numerical Weather Prediction, Proceedings of IUGG NWP Symposium}.
Meteorological Society of Japan, T. Matsuno, Ed., 831 pp : 247--253,

\item[] 
Williamson D.~L, J.~B.~Drake, J.~J.~Hack, R.~Jakob, P.~N.~Swarztrauber, 1992:
A standard test set for numerical approximation to the shallow water equations
in spherical geometry.
\emph{J.~Comp.~Phys.}, {\bf 102}, 211--224.

\end{itemize}


\end{document}